%% file: main.tex
\documentclass[runningheads]{llncs}

\input{preamble}

\def\orcidID#1{\href{http://orcid.org/#1}{\raisebox{-1.25pt}{\includegraphics{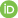}}}}
\usepackage{bbding}

\begin{document}
\title{Polar: An Algebraic Analyzer for (Probabilistic) Loops}
%
%
\author{
Marcel Moosbrugger\orcidID{0000-0002-2006-3741}\Envelope \and
Julian Müllner\orcidID{0009-0006-2909-2297} \and
Ezio Bartocci\orcidID{0009-0006-2909-2297} \and 
Laura Kov\'acs\orcidID{0000-0002-8299-2714}\Envelope
}
\authorrunning{Marcel Moosbrugger, Julian Müllner, Ezio Bartocci, and Laura Kov\'acs}
\institute{TU Wien, Vienna, Austria\\
\{firstname.lastname@tuwien.ac.at\}}
\maketitle              
\begin{abstract}
We present the \Polar{} framework for fully automating the analysis of classical and probabilistic loops using algebraic reasoning. The central theme in \Polar{} comes with handling algebraic recurrences that precisely capture the loop semantics. To this end, our work implements a variety of techniques to compute exact closed-forms of recurrences over higher-order moments of variables, infer invariants, and derive loop sensitivities with respect to unknown parameters. 
\Polar{} can analyze probabilistic loops containing if-statements, polynomial arithmetic, and common probability distributions.
By translating loop analysis into linear recurrence solving, \Polar{}  uses the derived closed-forms of recurrences to compute the strongest polynomial invariant or to infer parameter sensitivity. \Polar{} is both sound and complete within well-defined programming model restrictions. Lifting any of these restrictions results in significant hardness limits of computation. To overcome computational burdens for the sake of efficiency, \Polar{} also provides incomplete but sound techniques to compute moments of combinations of variables. 

\end{abstract}

\input{01-introduction}
\input{02-closed-forms}
\input{03-invariants}
\input{04-sensitivity}
\input{05-unsolvable-loops}
\input{06-related-work}
\input{07-conclusion}

\paragraph{Acknowledgements.} 
We thank  Daneshvar Amrollahi, George Kenison and Miroslav Stankovic for  valuable discussions and joint work related to Polar. We acknowledge  generous funding from the  ERC Consolidator Grant ARTIST 101002685, the WWTF 10.47379/ICT19018 grant ProbInG, 
the TU Wien Doctoral College SecInt,  
 and the Amazon Research Award 2023 QuAT. 

\newpage
\bibliographystyle{splncs04}
\bibliography{references}

\end{document}

%% file: preamble.tex
\usepackage{lmodern}
\usepackage[T1]{fontenc}
\usepackage[utf8]{inputenc}

\usepackage{amsmath}
\usepackage{amssymb}
\usepackage{mathtools}
\usepackage{microtype}
\usepackage[inline]{enumitem}
\usepackage{multirow}
\usepackage{booktabs}
\usepackage{subcaption}
\usepackage[usenames,dvipsnames,table]{xcolor}
\usepackage{nag}
\usepackage{hyperref}
\usepackage{tabto}
\usepackage{algpseudocode}
\usepackage{thmtools}
\usepackage{tcolorbox}
\usepackage{arydshln}
\usepackage{listings}
\usepackage{lstautogobble}
\usepackage{siunitx}
\usepackage{xargs}
\usepackage{pifont}
\usepackage{nicefrac}
\usepackage{syntax}
\usepackage{amsfonts}
\usepackage{graphicx}
\usepackage{algorithm}
\usepackage{mleftright}
\usepackage{nicematrix}
\usepackage{xspace}
\usepackage{fancyvrb}

\definecolor{hufflepuff}{RGB}{236,185,57}
\usepackage[colorinlistoftodos,prependcaption,obeyFinal]{todonotes}
\presetkeys{todonotes}{inline, color=hufflepuff!40, bordercolor=hufflepuff}{}

\definecolor{keywordsColor}{RGB}{97, 0, 71}
\definecolor{commentsColor}{RGB}{54, 54, 54}

\lstdefinelanguage{OurLanguage}{
	alsoletter={:,=, <, >, &, |},
	keywords={while, end, types, if, else, else:, and, or, not},
	morekeywords={=, <, >, <=, >=, ==, !=, &&, ||},
	basicstyle={\ttfamily\small\normalfont},
	keywordstyle={\color{keywordsColor}\ttfamily\bfseries},
	comment=[l]{\#},
	commentstyle={\color{commentsColor}\ttfamily},
	autogobble=true,
	mathescape=true
}
\lstdefinestyle{program}{basicstyle=\small\ttfamily,keywordstyle=\bfseries}

\newcommand{\Polar}{\textsc{Polar}}

\newcommand{\R}{\ensuremath{\mathbb{R}}\xspace}

\newcommand{\N}{\ensuremath{\mathbb{N}}\xspace}

\newcommand{\E}{\ensuremath{\mathbb{E}}}
\newcommand{\V}{\ensuremath{\mathbb{V}}}

\newcommand{\Inv}{\ensuremath{\mathcal{I}}\xspace}
\newcommand{\MInv}{\ensuremath{\mathbb{I}}\xspace}

\definecolor{algo1}{HTML}{4C0033}
\definecolor{algo2}{HTML}{01024E}

\lstdefinestyle{benchmark-code-style}{
    basicstyle=\ttfamily\footnotesize
}

\newcommand{\diffp}{\ensuremath{\partial_p}}

\makeatletter
\newcommand{\oset}[3][0ex]{%
  \mathrel{\mathop{#3}\limits^{
    \vbox to#1{\kern-2\ex@
    \hbox{$\scriptstyle#2$}\vss}}}}
\makeatother

\DeclarePairedDelimiter\abs{\lvert}{\rvert}
\makeatletter
\let\oldabs\abs
\def\abs{\@ifstar{\oldabs}{\oldabs*}}
\makeatother

\definecolor{cadmiumgreen}{rgb}{0.0, 0.42, 0.24}

\newcommand{\Q}{\ensuremath{\mathbb{Q}}\xspace}

%% file: 01-introduction.tex
\section{Introduction}

Probabilistic programs (PPs) extend classical programs by integrating the ability to draw from common probability distributions~\cite{Barthe2020,Kozen1985}.
This integration allows PPs to embed stochastic quantities within standard control flow, enabling the encoding of complex probability distributions.
As such, PPs constitute a unifying framework to model stochastic systems across various domains, such as machine learning~\cite{Ghahramani2015}, natural sciences~\cite{unsolvable}, cyber-physical systems~\cite{Chou2020,Selyunin2015}, security and privacy~\cite{Barthe2012,Barthe2012a}, or randomized algorithms~\cite{Motwani1995}.
In this context, probabilistic program loops can encode stochastic processes, modeling dynamics under uncertainty, such as dynamic Bayesian networks~\cite{Stankovic2022} or autonomous vehicles~\cite{KofnovMSBB22,Kofnov2024}.

Given the wide application range of PPs, automating the formal analysis of PPs is intrinsically hard. As pointed out by Joost-Pieter Katoen and his co-authors in~\cite{hark-aiming-2020},  
\lq\lq{}formal verification of probabilistic programs is strictly harder than for nonprobabilistic programs\rq\rq{}.
\emph{In this paper we address some of the verification challenges studied by Joost-Pieter Katoen in probabilistic programming} and present \Polar{}, a fully automatic tool for the automated analysis of nonprobabilistic and probabilistic loops. In the sequel, we refer to nonprobabilistic programs as classic programs. Our \Polar{} framework complements the many algorithmic approaches, see e.g. ~\cite{KatoenMMM10,Katoen2011}, Joost-Pieter Katoen has developed over the past years. In particular, \Polar{} combines algebraic reasoning with statistical methods, which we believe yields a rigorous methodology that Joost-Pieter Katoen will enjoy using and experimenting with.
\Polar{} is open-source and publicly available:

\begin{center}
\url{https://github.com/probing-lab/polar}\footnote{Throughout the paper, \Polar{} is used with \emph{version 1.0}.}
\end{center}

\paragraph{The \Polar{} methodology.} Formally verifying and analyzing PPs has received significant attention, due to their complexity introduced by the combination of randomness and deterministic control flow.
Deductive program calculi~\cite{Kozen1985,McIver2005}, such as the weakest preexpectation calculus, offer effective methods for reasoning about PPs.
Recently, various probabilistic program calculi have been proposed to address different program properties~\cite{BatzKKMV23,Aguirre2021,KlinkenbergBKKM20,Kaminski2016,KaminskiKMO18}.
The primary challenge in automating these calculi is handling program loops.
Similarly to classical program verification, verifying probabilistic loops requires invariant properties, which typically are provided manually. In contrast to these works, \Polar{} automates the analysis of classical and probabilistic loops without any user guidance, by relying on algebraic recurrence solving. Under certain restrictions on its programming model, \Polar{} computes closed-forms of loop recurrences over moments of loop variables. These closed-forms are further used to derive (strongest) loop invariants and loop sensitivities of variable moments with respect to unknown parameters.

\begin{figure}[t]
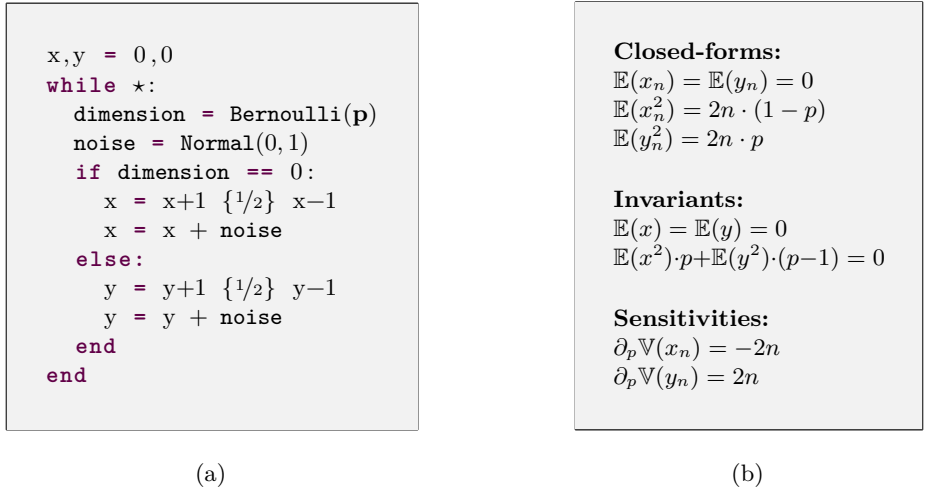

    \centering
    \begin{subfigure}[b]{0.45\textwidth}
        \begin{tcolorbox}[width=\linewidth,boxrule=0.5pt,arc=0pt,auto outer arc]
        \begin{lstlisting}
            x,y = 0,0
            while $\star$:
              $\texttt{dimension}$ = $\texttt{Bernoulli}(\textbf{p})$
              $\texttt{noise}$ = $\texttt{Normal}(0,1)$
              if $\texttt{dimension}$ == 0:
                x = x+1 {$\nicefrac{1}{2}$} x-1
                x = x + $\texttt{noise}$
              else:
                y = y+1 {$\nicefrac{1}{2}$} y-1
                y = y + $\texttt{noise}$
              end
            end
        \end{lstlisting}
        \end{tcolorbox}
        \caption{}
        \label{fig:intro:loop}
    \end{subfigure}\hfill
    \begin{subfigure}[b]{0.38\textwidth}
        \begin{tcolorbox}[width=\linewidth,boxrule=0.5pt,arc=0pt,auto outer arc]

        \vspace{0.7em}
        
        \textbf{Closed-forms:} \\
        $\E(x_n) = \E(y_n) = 0$ \\
        $\E(x^2_n) = 2n \cdot (1-p)$ \\
        $\E(y^2_n) = 2n \cdot p$

        \bigskip

        \textbf{Invariants:} \\
        $\E(x) = \E(y) = 0$ \\
        $\E(x^2) \cdot p + \E(y^2) \cdot (p - 1) = 0$

        \bigskip

        \textbf{Sensitivities:} \\
        $\diffp \V(x_n) = -2n$ \\
        $\diffp \V(y_n) = 2n$

        \vspace{0.7em}
        \end{tcolorbox}
        \caption{}
        \label{fig:intro:data}
    \end{subfigure}
    \caption{A probabilistic loop modeling a two-dimensional random walk with Gaussian noise and parameter \texttt{p}. Fig.~\ref{fig:intro:data} depicts information about the loop as computed by \Polar{}.}
    \label{fig:intro}
\end{figure}

\paragraph{Algebraic loop analysis via \Polar{}.} The central feature of \Polar{} comes with its capability to extract algebraic recurrences that precisely describe statistical moments of loop variables, such as expected values or even higher moments~\cite{polar}.
Specifically, these recurrences are linear recurrences with constant coefficients, so-called \emph{C-finite recurrences}~\cite{Kauers2011}.
In our setting, \Polar{} computes closed-forms of such recurrences as being functions representing moments of loop variables parameterized by the number of loop iterations $n$. 
Figure~\ref{fig:intro:loop} illustrates a probabilistic loop modeling a two-dimensional random walk with Gaussian noise and an unknown parameter \texttt{p}.
The closed-forms, automatically computed by \Polar{} and shown in Figure~\ref{fig:intro:data}, provide the expected values and second moments of the variables \texttt{x} and \texttt{y} for any loop iteration $n$ and value of \texttt{p}.

Two central concepts in the analysis of probabilistic loops are \emph{loop invariants} and \emph{sensitivities}.
For moments of program variables, invariants are properties of the moments that are valid throughout a loop's execution.
Sensitivities quantify how changes in unknown parameters affect variable moments.
\Polar{} leverages the computed closed-forms of loop recurrences to automatically infer (strongest) loop invariants and quantify the sensitivity of variable moments on unknown parameters.
Figure~\ref{fig:intro:data} shows the \emph{strongest polynomial invariant} among the first two moments of the variables \texttt{x} and \texttt{y} for the loop depicted in Figure~\ref{fig:intro:loop}, as well as the sensitivities of their variances with respect to the parameter \texttt{p}.
For classical loops, \Polar{}  computes \emph{the} strongest polynomial invariant of the loop.

Loops that can be analyzed by \Polar{} support if-statements, symbolic constants, polynomial arithmetic, constant probabilistic choice, and draws from common probability distributions.
To ensure computability, \Polar{} imposes certain restrictions on its input programs (see Section~\ref{sec:closed-forms:restrictions}): 
besides enforcing probabilities to be constant, all variables in if-conditions must only assume finitely many values and all cyclic variable dependencies must be linear.
Interestingly, none of the restrictions can be lifted without encountering significant hardness boundaries~\cite{polar,MullnerMK24}.
Despite these hardness results, \Polar{} implements an incomplete, but sound, method to compute closed-forms for combinations of program variables, even for certain loops that violate the computability restrictions~\cite{unsolvableloopanalysis}.


\paragraph{Contributions.} This paper serves as the first comprehensive tutorial on \Polar{}, discussing what \Polar{} can do and how it can be used. 
We present the main features and techniques implemented in \Polar, designed for the automated analysis of both classical and probabilistic loops, as follows. 
\begin{itemize}
    \item We describe the use of C-finite recurrences for computing closed-forms for (moments of) loop variables \cite{polar}. We also detail the restrictions on input programs that \Polar{} enforces to guarantee computability (Section~\ref{sec:closed-forms}).

    \item We provide a comprehensive overview of how \Polar{} uses exponential polynomial closed-forms to compute the strongest polynomial invariant for classical and probabilistic loops (Section~\ref{sec:invariants}).

    \item We explain how \Polar{} can be used to compute the sensitivities of variable moments with respect to unknown model parameters \cite{MoosbruggerMK23} (Section~\ref{sec:sensitivity}).

    \item We illustrate the incomplete, but  sound techniques that \Polar{} employs to analyze loops that violate its computability restrictions \cite{unsolvableloopanalysis} (Section~\ref{sec:unsolvable}).
\end{itemize}

%% file: 02-closed-forms.tex
\section{Computing Closed-Form Formulas}\label{sec:closed-forms}

\begin{figure}[t]
    {
    \begin{minipage}{\linewidth}
    \scriptsize
    $\mathit{lop} \in \{ \mathit{and}, \mathit{or} \}$,
    $\mathit{cop} \in \{ =, \neq, <, >, \geq, \leq \}$,
    $\mathit{Dist} \in \{ \mathit{Bernoulli},  \mathit{Beta}, \mathit{Categorical}, \mathit{DiscreteUniform},$ \\ $\mathit{Exponential}, \mathit{Gamma}, \mathit{Laplace}, \mathit{Normal},  \mathit{TruncNormal}, \mathit{Uniform} \}$
    \begin{grammar}
    	<sym> ::= "a" | "b" | $\dots$ <var> ::= "x" | "y" | $\dots$
    	
    	<const> ::= $r \in \R$ | <sym> | <const> ( "+" | "*" | "/" ) <const>
    	
    	<poly> ::= <const> | <var> | <poly> ("+" | "-" | "*") <poly> | <poly>"**n"
    	
    	<assign> ::= <var> "=" <assign\_right> | <var> "," <assign> "," <assign\_right>
    	
    	<categorical> ::= <poly> ("\{"<const>"\}" <poly>)* ["\{"<const>"\}"]
    	
    	<assign\_right> ::= <categorical> | Dist"("<poly>$^*$")"
    	
    	<bexpr> ::= "true" ($\star$) | "false" | <poly> <cop> <poly> | "not" <bexpr> | <bexpr> <lop> <bexpr>
    	
    	<ifstmt> ::= "if" <bexpr>":" <statems> ("else if" <bexpr>":" <statems>)$^*$ ["else:" <statems>] "end"
    	
    	<statem> ::= <assign> | <ifstmt> \quad \quad \quad <statems> ::= <statem>$^+$
    	
    	<loop> ::= <statem>* "while" <bexpr> ":" <statems> "end"
    \end{grammar}
    \end{minipage}
    }
    \caption{Grammar describing the syntax of the input programs to \Polar{} \cite{polar}.}
    \label{fig:syntax}
\end{figure}

Assignments in program loops capture relations between the values of
variables in previous loop iterations and variable values in the current iteration.
Likewise, algebraic recurrence relations describe values of sequences in terms of previous values.
This similarity renders algebraic recurrence solving as a natural tool for the analysis of loops.
Linear recurrences with constant coefficients, also referred to as \emph{C-finite recurrences}, are particularly suitable for automation:
the sequences described by C-finite recurrences are always expressible
with computable closed-form formulas~\cite{Kauers2011}.

A core functionality of \Polar{}  comes with \emph{modeling loops by
recurrence equations}, which in turn enables \Polar{} to compute closed-forms of
recurrences  over (i)   loop variable values in classical, non-probabilistic loops; (ii) 
as well as expected values and higher moments of loop variables in probabilistic loops.
These closed-forms are further used to infer loop invariants
or automate sensitivity analysis over loop parameters (see Sections~\ref{sec:invariants}--\ref{sec:sensitivity}).

\begin{figure}[t]
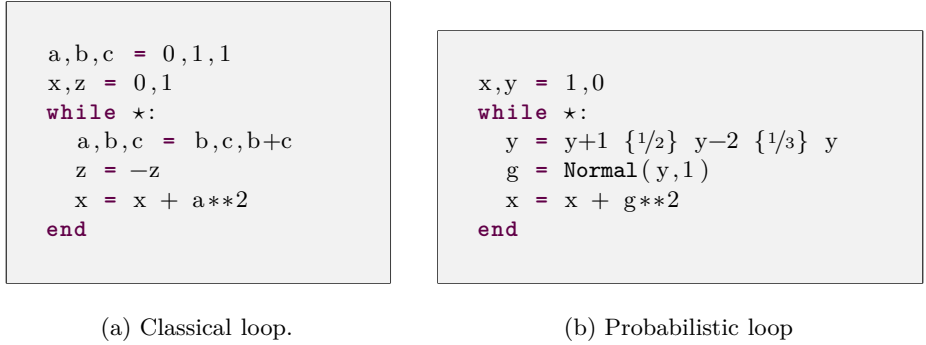

    \centering
    \begin{subfigure}[b]{0.42\textwidth}
        \begin{tcolorbox}[width=\linewidth,boxrule=0.5pt,arc=0pt,auto outer arc]
        \begin{lstlisting}
            a,b,c = 0,1,1
            x,z = 0,1
            while $\star$:
              a,b,c = b,c,b+c
              z = -z
              x = x + a**2
            end
        \end{lstlisting}
        \end{tcolorbox}
        \caption{Classical loop.}
        \label{fig:closed-forms:fib}
    \end{subfigure}\hfill
    \begin{subfigure}[b]{0.53\textwidth}
        \begin{tcolorbox}[width=\linewidth,boxrule=0.5pt,arc=0pt,auto outer arc]
        \begin{lstlisting}
            x,y = 1,0
            while $\star$:
              y = y+1 {$\nicefrac{1}{2}$} y-2 {$\nicefrac{1}{3}$} y
              g = $\texttt{Normal}$(y,1)
              x = x + g**2
            end
        \end{lstlisting}
        \end{tcolorbox}
        \caption{Probabilistic loop}
        \label{fig:closed-forms:prob}
    \end{subfigure}
    \caption{Two loops for which closed-forms of loop variables (moments) are computable with \Polar. Here, ** denotes the power operator.}
    \label{fig:closed-forms}
  \end{figure}

\paragraph{The \Polar{} programming model.} \Polar{}'s input programs are non-nested while-loops with standard program constructs.
Figure~\ref{fig:syntax} depicts the grammar defining the input programs.
\Polar{} supports symbolic constants, probabilistic choice between different expressions, and drawing from probability distributions with existing moments.
For the probabilistic loop in Figure~\ref{fig:closed-forms:prob}, the assignment of $y$ contains a probabilistic choice between three expressions which evaluates to $y+1$ with probability $\nicefrac{1}{2}$, $y-2$ with probability \nicefrac{1}{3} and $y$ with the remaining probability of $\nicefrac{1}{6}$.
Conditioning is a prevalent feature in probabilistic program languages and is currently not supported by \Polar{}.

The following example illustrates C-finite recurrences and their use
in the automated analysis of classical loops. Here and in sequel, by a
classical loop we mean a non-probabilistic loop.

\begin{example}[Fibonacci loop]\label{ex:fib-loop}
The classical loop in Figure~\ref{fig:closed-forms:fib} iterates over 
Fibonacci numbers using the program variables $a$, $b$, and $c$.
Moreover, the program variable $z$ toggles between $1$ and $-1$.
The value of $x$ after $n$ loop iterations corresponds to the sum of squares of the first $n$ Fibonacci numbers.
The value sequences of the program variables $a$, $b$, $c$,  and $z$ are
fully described by the following system of C-finite recurrences
together with the initial values of loop variables; here, the initial values
of loop variables are already substituted in the recurrences. 
\begin{gather}\label{eq:fib-rec}
\begin{aligned}
&a_{n+1} = b_n &\hspace{5em} b_{n+1} = c_n \\
&c_{n+1} = b_n + c_n &z_{n+1} = -z_n
\end{aligned}
\end{gather}
where $a_{n}$ denote the value of $a$ at the $n$th loop iterations;
similarly for the other variables/sequences.
The  system~\eqref{eq:fib-rec} of recurrences can automatically be constructed from the assignments in the loop body and succinctly expressed using matrices and vectors:
\begin{align}\label{eq:cf:matrix}
\begin{pNiceMatrix}a \\ b \\ c \\ z \end{pNiceMatrix}_{n+1}
= 
\underbrace{\begin{pNiceMatrix}[columns-width=auto]
0 & 1 & 0 & 0 \\
0 & 0 & 1 & 0 \\
0 & 1 & 1 & 0 \\
0 & 0 & 0 & -1
\end{pNiceMatrix}}_{M}
\cdot
\begin{pNiceMatrix}a \\ b \\ c \\ z \end{pNiceMatrix}_{n}
\end{align}
Closed-forms of each component of the matrix
system~\eqref{eq:cf:matrix} are expressed as an \emph{exponential polynomial}, that is, a sum of polynomials and exponentials.
More precisely, for every variable $\alpha$ of the system, we have 
$$\alpha_n = \sum_{\lambda} p_{\lambda}(n) \lambda^n,$$
where the $\lambda$'s are the eigenvalues of the matrix $M$ and every $p_{\lambda}(n)$ is a polynomial depending on $\lambda$ and the initial value of $\alpha$.
Such a closed-form representation is not specific to this example but
holds in general for any system of C-finite recurrences~\cite{Kauers2011}.
For the loop in Figure~\ref{fig:closed-forms:fib}, we obtain the
following closed-forms: 
\begin{gather*}
F_n = \frac{1}{\sqrt{5}} \left( \frac{1 + \sqrt{5}}{2} \right)^n - \frac{1}{\sqrt{5}} \left( \frac{1 - \sqrt{5}}{2} \right)^n \\
\begin{alignedat}{2}
a_n &= F_n &\hspace{5em} b_n &= F_{n+1} \\
c_n &= F_{n+2} & z_n &= (-1)^n
\end{alignedat}
\end{gather*}
where $F_n$ is \emph{Binet's formula} for Fibonacci numbers.

So far, the program variable $x$ of Figure~\ref{fig:closed-forms:fib} was omitted in this example.
Although the assignment of $x$ in the loop body contains a non-linear term, its values can still be described by C-finite recurrences.
The main idea is to construct linear recurrences for monomials in program variables.
To fully capture the values of $x$, the system \eqref{eq:fib-rec} can be extended by the following recurrences:
\begin{gather}\label{eq:cf:x}
  \begin{aligned}
&x_{n+1} = x_n + b^2_n
&b^2_{n+1} = c^2_n \\
&c^2_{n+1} = b^2_n + 2cb_n + c^2_n
&cb_{n+1} = cb_n + c^2_n
  \end{aligned}
  \end{gather}
The resulting system~\eqref{eq:cf:x} is comprised of C-finite recurrences over monomials in program variables and yields the closed-form for the program variable $x$.
\end{example}

\subsection{Closed-Forms for Probabilistic Loops}

Our tool \Polar{} implements the computation of closed-forms for classical loops and can also analyze probabilistic loops.
For probabilistic loops, the exact values of program variables are inherently stochastic and hence do not follow classical recurrences.
Nevertheless, expected values and higher moments of program variables can obey C-finite recurrences, referred to as \emph{moment recurrences} \cite{Bartocci2019,moosbrugger2020automated}.

\begin{example}[Closed-forms for Moments]\label{ex:pp-closed-forms}
The probabilistic loop in Figure~\ref{fig:closed-forms:prob} contains assignments with probabilistic choice and draws from probability distributions.
Hence, the values of the loop variables do not follow classical
algebraic recurrences.
However, using the expected value operator $\E$, we can devise recurrences
for moments of program variables, as illustrated next.

Due to the assignment of $x$ in Figure~\ref{fig:closed-forms:prob}, we have $\E(x_{n+1}) = \E(x_n) + \E(g^2_{n+1})$.
The assignment of $g$ contains a distribution with a state-dependent
parameter,  which can be resolved using the identity $\texttt{Normal}(y,1) = y + \texttt{Normal}(0,1)$.
Hence, we construct the following recurrences for the first two moments of $g$:
\begin{flalign*}
    \E(g_{n+1}) &= \E(y_{n+1}) + \E(\texttt{Normal}(0,1)) = \E(y_{n+1}) \\
    \E(g^2_{n+1}) &= \E(y^2_{n+1}) +  \E(2y_{n+1}\texttt{N}(0,1)) + \E(\texttt{N}(0,1)^2) \\
    &= \E(y^2_{n+1}) + 1
\end{flalign*}

Similarly, the recurrences for the first two moments of the  loop 
variable $y$ are: 
\begin{flalign*}
    \E(y_{n+1}) &= \frac{\E(y_n + 1)}{2} + \frac{\E(y_n - 1)}{3} + \frac{\E(y_n)}{6} = \E(y_n) - \frac{1}{6} \\
    \E(y^2_{n+1}) &= \frac{\E((y_n + 1)^2)}{2} + \frac{\E((y_n - 1)^2)}{3} + \frac{\E(y^2_n)}{6} \\
    &= \E(y^2_n) - \frac{1}{3}\E(y_n) + \frac{11}{6}
\end{flalign*}

By combining the moment recurrences of $y$ and $g$ together with the
moment recurrences of $x$,
we obtain a system of C-finite recurrences over expected values of
monomials in the program variables of
Figure~\ref{fig:closed-forms:prob}. Solving this recurrence system, \Polar{} infers the following closed-forms for the loop variable moments:
\begin{flalign*}
\E(y_n) &= -\frac{n}{6} \qquad \E(y^2_n) = \frac{n^2 + 65n}{36} \\
\E(x_n) &= \frac{n^3}{256} + \frac{133 n^2}{256} + \frac{205 n}{128} + 1
\end{flalign*}
\end{example}

As Examples~\ref{ex:fib-loop} and \ref{ex:pp-closed-forms} suggest, the system of recurrences can often be constructed by applying the expected value operator on assignments together with bottom-up substitution of program variables by their assignments as they appear in the loop body.
Throughout the process, properties such as the linearity of expectation are used to simplify the resulting recurrences.
The treatment of if-statements in the construction of recurrences is non-trivial.
We refer to \cite{polar} for more details on constructing moment recurrences, especially regarding the support of if-statements.

\subsection{Computing Closed-Forms with \Polar{}}

\Polar{} utilizes
moment recurrences to automatically compute closed-forms for raw
moments such as $\E(x^k_n)$,  for a fixed $k \in \N$.
These closed-forms are parameterized by the number of loop iterations
$n \in \N$ and can also be computed for mixed moments, such as $\E(xy_n)$.
Moreover, central moments can be computed using raw moments.
For instance, the variance of a variable $x$ is determined by its first two raw moments: $\V(x) = \E(x^2) - \E(x)^2$.
\Polar{} provides the following command-line interface for closed-form computation, where \texttt{BENCHMARK} is a path to the input program to analyze.
\begin{verbatim}
    python polar.py BENCHMARK --goals "GOAL1", ..., "GOALk"
\end{verbatim}

\Polar{} derives a closed-form form for each  goal from {\tt GOAL1,
  ..., GOALk},
passed through the \texttt{goals} parameter.
Each goal
from {\tt GOAL1,
  ..., GOALk},
represents an expected value, a central moment, or a cumulant for some
monomial $M$ in program variables, and has one of the following forms:
\begin{itemize}
\item \texttt{E(M)} for the expected value of $M$;
 \item \texttt{cd(M)} for the central moment of degree $\texttt{d}$ of
   $M$, where $\texttt{d} \in \N$;
   \item \texttt{kd(M)} for the cumulant of degree $\texttt{d}$ of
     $M$, with $\texttt{d} \in \N$.
     \end{itemize}
For guarded loops, \Polar{} can compute moments of program variables after termination of the loop by additionally passing the flag \texttt{after_loop}.

\subsection{Restrictions for Computability}\label{sec:closed-forms:restrictions}

The class of input programs defined by the grammar in Figure~\ref{fig:syntax} is Turing-complete.
Hence, closed-forms are uncomputable in general.
Nevertheless, with some further restrictions (R) on the input programs as
listed  below,
\Polar{} guarantees computability~\cite{moosbrugger2020automated}.

\paragraph{R1 -- Constant probabilities.}
The first restriction is that probabilities in probabilistic choice
constructs, as well as distribution parameters, must be constant, which means, they must not change between loop iterations.
Similarly to Example~\ref{ex:pp-closed-forms}, some non-constant distribution parameters are supported by utilizing specific identities to eliminate the dependency of parameters on the program state.

\paragraph{R2 -- Finite guards.}
Program variables occurring in if-statements and/or the loop guard must only assume finitely many values through the loop's execution.
Intuitively, this is the restriction that prevents Turing-completeness of the input programs.

\paragraph{R3 -- Acyclic Non-Linearity.}
\Polar{}'s program model allows polynomial arithmetic.
However, if a program variable $x$ depends non-linearly on $y$, then $y$ must not depend on $x$.

\medskip

With these three restrictions {\it R1--R3}, the closed-form of any moment of program variables is computable by \Polar{}.
Restriction {\it R2} is necessary because, without it, the program model is Turing-complete.
However, also restrictions {\it R1 } and {\it R3} cannot be lifted
without facing serious hardness boundaries on computability (cf. Section~\ref{sec:unsolvable}).

%% file: 03-invariants.tex
\section{Computing Invariants}\label{sec:invariants}

Loop invariants are properties that hold throughout the execution of a loop and are essential for verifying the safety and correctness of programs containing loops.
In general, invariants strong enough for verifying a given safety property are uncomputable.
Hence, existing works either resort to incomplete methods or impose restrictions on the class of supported loops and invariants.
Polynomial invariants are given by polynomial equations in program variables.
Every polynomial equation $P = Q$ can trivially be transformed to $P - Q = 0$.
Therefore, it is sufficient to only consider polynomial invariants of the form $P = 0$, where $P$ is a polynomial in program variables.
Polynomial invariants constitute a relatively tractable, yet expressive class of invariants.

\begin{example}\label{ex:fibonacci-ideal}
For the loop in Figure~\ref{fig:closed-forms:fib}, the variables $a$, $b$, and $c$ are three consecutive Fibonacci numbers, $z$ toggles between $1$ and $-1$, and the variable $x$ sums the squares of $a$.
The following equalities are polynomial invariants of the loop and correspond to well-known identities involving Fibonacci numbers:
\begin{flalign*}
a + b - c = 0 && \text{(Fibonacci definition)} \\
z - b^2 - bc - c^2 = 0 && \text{(Cassini's identity)} \\
b^4 + 2b^3c - b^2c^2 - 2bc^3 + c^4 - 1 = 0 && \text{(Quartic identity)}\\
b^2 - bc + x = 0 && \text{(Sum of squares identity)}
\end{flalign*}
\end{example}

The \emph{strongest polynomial invariant} of a loop is the set \Inv of \emph{all} polynomials $P$ in program variables such that $P = 0$ is an invariant.
The fact that \Inv generally contains infinitely many elements is mitigated by its well-behaved structure:
the strongest polynomial invariant forms an \emph{ideal}.
This means \Inv is a group and, for every $P \in \Inv$ and \emph{every} polynomial in program variables $Q$, we have $P \cdot Q \in \Inv$.
For a given loop, \Inv is also called the \emph{invariant ideal} of the loop.
Crucially, Hilbert's basis theorem states that every polynomial ideal is finitely generated~\cite{CoxLittleOshea97}.
Hence, \Inv can be fully characterized by listing a finite number of polynomials.
In fact, the four polynomial invariants in Example~\ref{ex:fibonacci-ideal} form a basis and generate the strongest polynomial invariant of the Fibonacci loop from Figure~\ref{fig:closed-forms:fib}.

When considering polynomial invariants in probabilistic loops, we note
that in general $\E(x^k) \neq \E(x)^k$.
Intuitively speaking, this is the main reason why the set of all polynomial invariants among all moments of program variables is not a polynomial ideal and is not finitely generated.
A solution to infer
polynomial invariants among moments of probabilistic loop variables is to fix a finite set of moments and only consider polynomial invariants over these moments~\cite{MullnerMK24}.
For instance, for the probabilistic loop of  Figure~\ref{fig:closed-forms:prob} the following single invariant is a basis for all polynomial invariants among $\E(x)$ and $\E(y)$:
\begin{flalign*}
    \E(x) + 2\E(y)^3 - 33\E(y)^2 + \frac{103}{9}\E(y) - 1 = 0
\end{flalign*}

For a given loop and a finite set $M$ of moments of program variables, the strongest polynomial invariant is the set \MInv of all polynomials $P$ in $M$ such that $P = 0$ is an invariant.
For instance, if $M = \{ \E(x), \E(y) \}$, then $\MInv$ contains all polynomial invariants among the expected values of the program variables $x$ and $y$.
The set \MInv is also referred to as the \emph{moment invariant ideal} with respect to $M$ \cite{MullnerMK24}.
Importantly, moment invariant ideals are a proper generalization of
classical polynomial invariants: for every program variable $x$ of any classical loop we have $\E(x) = x$.
Hence, for classical loops, the moment invariant ideal with respect to
the first moments of loop variables corresponds to the classical invariant ideal.

\subsection{From Closed-Forms to Invariant Ideals}

Our tool \Polar{} can compute a basis for the invariant ideal of any classical loop and any moment invariant ideal of probabilistic loops adhering to the computability restrictions from Section~\ref{sec:closed-forms:restrictions}.
It does so, by first computing the closed-forms of all program variables (or a fixed set of moments for probabilistic loops) as exponential polynomials parameterized by the number of loop iterations $n$ (cf. Section~\ref{sec:closed-forms}).
Second, \Polar{} uses existing techniques to compute all polynomial relations among the closed-forms that hold for all $n \in \N$.
In the remainder of this section, we sketch the main ideas of how \Polar{} computes all polynomial relations among exponential polynomial closed-forms.

\subsubsection{Polynomial Closed-Forms.}
\Polar{} uses C-finite recurrences to compute closed-forms for (moments of) loop variables which, in general, are exponential polynomials.
However, if all closed-forms are standard polynomials, such as in Example~\ref{ex:pp-closed-forms}, the computation of bases for (moment) invariant ideals simplifies significantly.
We illustrate the basis computation for this special case with the following example.

\begin{example}\label{ex:ideal-poly}
Assume a classical loop with variables $x$, $y$ and polynomial closed-forms $x_n = n^2 - 1$ and $y_n = n^3 + n$.
The ideal $\hat{I}$ generated by $x - n^2 - 1$ and $y - n^3 - n$ is an ideal in the polynomial ring $\Q[x,y,n]$ and subsumes the loop's invariant ideal \Inv.
However, $\hat{I}$ references $n$ which is not a variable of the loop.
Intuitively, $n$ needs to be eliminated from $\hat{I}$ to compute \Inv.
More precisely, $\Inv = \hat{I} \cap \Q[x,y]$.
A basis for $\hat{I} \cap \Q[x,y]$, and hence \Inv, can be inferred
using Gr\"obner bases computation via Buchberger's
algorithm~\cite{Buchberger-thesis}. In this respect,
the first step is to compute a Gr\"obner basis for $\hat{I}$ with
respect to a lexicographic monomial order such that $n > x,y$; as a
result, we obtain the  Gr\"obner basis: 
\begin{align*}
&n^2 - x - 1 &\hspace{1em} &nx + 2n - y \\
&ny - x^2 - 3x - 2 & &x^3 + 5x^2 + 8x - y^2 + 4
\end{align*}
Then, a basis for $\hat{I} \cap \Q[x,y]$, and hence \Inv, is given by
all above basis elements that do not contain $n$.
Therefore, the invariant ideal of $\Inv$ is generated by the single invariant $x^3 + 5x^2 + 8x - y^2 + 4 = 0$.
This technique to eliminate variables from polynomial ideals is
applicable in general: (i) compute a Gr\"obner basis with respect to a
lexicographic order such that the variables to eliminate have higher
precedence,  and (ii) in the resulting basis only retain elements
that do not contain ``unwanted'' variables, that is perform variable elimination~\cite{CoxLittleOshea97}.
\end{example}

\subsubsection{General Closed-Forms as Exponential Polynomials.}
Generally, closed-forms of C-finite recurrences contain polynomials \emph{and} exponentials.
Gr\"obner bases and Buchberger's algorithm are inherently tied to polynomial arithmetic and hence do not suffice for computing invariant ideals for all loops supported by \Polar{}.
The main idea for treating exponential terms is to replace them with
fresh variables,  transforming the closed-forms into polynomials.
There are no algebraic relations between any exponential $b^n$ and polynomial $p(n)$ that hold for all $n \in \N$.
On the other hand, there can be algebraic relations between multiple exponential terms.
However, any information about the exponential terms is removed when replacing them with fresh variables.
For this reason, the \emph{multiplicative relation}s between the exponentials are calculated separately and added to the basis containing the invariant ideal.
Then, the loop counter $n$ as well as the variables introduced for the exponentials are eliminated, as in Example~\ref{ex:ideal-poly}, resulting in a basis for the invariant ideal.

\begin{example}\label{ex:exp}
Assume a classical loop with variables $x$, $y$ and closed-forms $x_n = n2^n$ and $y_n = n^2 4^n$.
Because the closed-forms contain exponentials, we associate $2^n$ and
$4^n$ with the fresh variables $a$ and $b$, respectively; that is,
$a=2^n$ and $b=4^n$. 
The equation $a^2b - 1 = 0$ characterizes the multiplicative relations between $2^n$ and $4^n$.
The ideal $\hat{I}$ generated by $x - na$, $y - n^2b$, and $a^2b - 1$ is an ideal in the polynomial ring $\Q[x,y,n,a,b]$ and subsumes the loops invariant ideal.
In contrast to Example~\ref{ex:ideal-poly}, we now eliminate not only $n$ but also $a$ and $b$, resulting in the invariant ideal generated by the single polynomial $x^2 - y$.
\end{example}

Usually, the multiplicative relationships among exponentials are less evident than in Example~\ref{ex:exp} and need to be computed automatically.
The exponential bases occurring in the closed-forms of C-finite recurrences are algebraic numbers.
Existing algorithms can be used to compute the multiplicative relations among general algebraic numbers~\cite{kauers-thesis,KauersNP23}.
The algorithms are non-trivial and involve concepts from abstract algebra.
If all exponential bases are rational, \Polar{} uses a considerably simpler approach.
This simpler method involves factorization and solving a system of linear Diophantine equations which is particularly suitable if the numerators and denominators of the exponential bases are small.
We illustrate this approach with the following example.

\begin{example}
The multiplicative relations between the rational numbers $2$, $\nicefrac{1}{4}$, and $\nicefrac{1}{6}$ are given by the natural solutions of the following equation:
\begin{equation*}
2^x \cdot  \left( \frac{1}{4} \right)^y \cdot \left( \frac{1}{6} \right)^z = 1
\Longleftrightarrow
\left( \frac{2^x}{1} \right) \cdot \left( \frac{1}{2^y 2^y} \right) \left( \frac{1}{2^z 3^z} \right) = 1
\end{equation*}
For any solution, the multiplicity of every prime factor in the enumerator must be equal to the multiplicity in the denominator.
More precisely, a triple $(x,y,z)$ is a solution if and only if it is a solution to the following linear system:
\begin{align*}
x - 2y - z &= 0 \\
-z &= 0
\end{align*}
The set of all natural solutions of the linear system forms is characterized by a finite number of minimal solutions computable by standard means.
For this example, a basis is given by the single triple $(2,1,0)$.
The approach from this example generalizes to arbitrary rational numbers with one additional step.
Potential occurrences of the factor $-1$ are treated separately by adding a single equation to the linear system, ensuring that the multiplicity of $-1$ is even.
\end{example}

\begin{figure}[t]
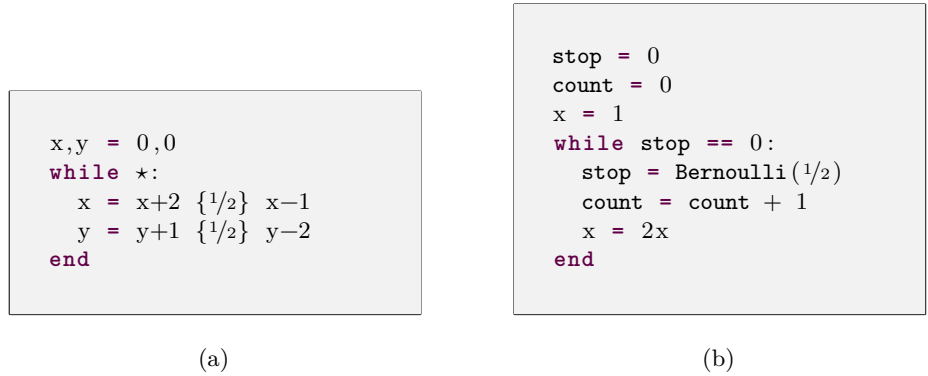

    \centering
    \begin{subfigure}[b]{0.45\textwidth}
        \begin{tcolorbox}[width=\linewidth,boxrule=0.5pt,arc=0pt,auto outer arc]
        \begin{lstlisting}
            x,y = 0,0
            while $\star$:
              x = x+2 {$\nicefrac{1}{2}$} x-1
              y = y+1 {$\nicefrac{1}{2}$} y-2
            end
        \end{lstlisting}
        \end{tcolorbox}
        \caption{}
        \label{fig:invariants:rw}
    \end{subfigure}\hfill
    \begin{subfigure}[b]{0.45\textwidth}
        \begin{tcolorbox}[width=\linewidth,boxrule=0.5pt,arc=0pt,auto outer arc]
        \begin{lstlisting}
            $\texttt{stop}$ = 0
            $\texttt{count}$ = 0
            x = 1
            while $\texttt{stop}$ == 0:
              $\texttt{stop}$ = $\texttt{Bernoulli}$($\nicefrac{1}{2}$)
              $\texttt{count}$ = $\texttt{count}$ + 1
              x = 2x
            end
        \end{lstlisting}
        \end{tcolorbox}
        \caption{}
        \label{fig:invariants:geo}
    \end{subfigure}
    \caption{Two probabilistic loops with computable moment invariant ideals.}
    \label{fig:invariants}
\end{figure}

\subsection{Computing Invariants with \Polar{}} 
\Polar{} implements (i)  the algorithm from \cite{KauersNP23} to
compute the multiplicative relations among exponentials with general
algebraic number bases, and (ii) a simpler method for rational bases.
Together with the closed-form computation introduced in
Section~\ref{sec:closed-forms} and Gr\"obner bases manipulations, \Polar{}
infers (moment) invariant ideals fully automatically.
Computability of bases for (moment) invariant ideals is guaranteed for
any (probabilistic) loop adhering to the restrictions from
Section~\ref{sec:closed-forms:restrictions}~\cite{moosbrugger2020automated,MullnerMK24}.

\begin{example}
Figure~\ref{fig:invariants} contains two examples of probabilistic
loops. For these loops, \Polar{}  computes moment invariant ideals for
any set of program variable moments.
The loop in Figure~\ref{fig:invariants:rw} models two asymmetric random walks, both starting at the origin.
The moment invariant ideal for all first- and second-order moments is
characterized by the following invariants computed by \Polar{}: 
\begin{gather}\label{eq:rw-basis}
\begin{aligned}
\E(x^2) - \E(y^2) &= 0 \\
\E(xy)^2 + 2\E(xy)\E(y^2) + \frac{81}{4}\E(xy) + \E(y^2)^2 &= 0 \\
\frac{2}{9}\E(xy) + \E(y) + \frac{2}{9}\E(y^2) &= 0 \\
\E(x) - \frac{2}{9}\E(xy) - \frac{2}{9}\E(y^2) &= 0
\end{aligned}
\end{gather}
The moment invariant ideal contains all polynomial relations among the moments $\E(x_n)$, $\E(y_n)$, $\E(x^2_n)$, and $\E(xy_n)$ that are true after all number of loop iterations $n \in \N$.
We omit the index $n$ from moments in invariants to emphasize that the invariants are true for all $n \in \N$.
The ideal can be \lq\lq queried\rq\rq{} for invariants using the basis \eqref{eq:rw-basis}.
For instance, it can be automatically checked that $\E(xy) =
\E(x)\E(y)$ is an invariant, showing that the two random walks of
Figure~\ref{fig:invariants:rw}  are uncorrelated.

The probabilistic loop in Figure~\ref{fig:invariants:geo} contains a loop guard.
Because the single variable \texttt{stop} occurring in the loop guard can only assume the values $0$ and $1$, the loop is supported by \Polar{}.
The following single invariant is a basis for the moment invariant
ideal with respect to the expected values of the program variables: 
\begin{equation*}
    -\E(\texttt{count}) + 2\E(\texttt{stop}) = 0
\end{equation*}
Upon termination of the loop, we have $\texttt{stop} = 1$ and hence $\E(\texttt{stop}) = 1$.
Combining this fact obtained from the loop guard with the invariant from the basis results in the expected value of $\texttt{count}$ after termination, that is $\E(\texttt{count}) = 2$.
There is a plethora of calculi, such as the weakest-preexpectation calculus~\cite{McIver2005,Barthe2020}, that enable reasoning about expected values after termination.
However, regarding automation, these calculi require a manually provided loop invariant.
Investigating the connection between these calculi and moment
invariant ideals is an interesting direction for future research.
\end{example}

In summary, \Polar{} provides the functionality of computing (moment) invariant ideals through its command-line interface and the parameter \texttt{invariants}:

\begin{Verbatim}[commandchars=\\\{\}]
    python polar.py BENCHMARK --goals GOALS \textbf{--invariants}
\end{Verbatim}

With this command, \Polar{} first computes the closed-forms for all
goals passed through the \texttt{goals} parameter and then uses its
 approach described in this section to compute all algebraic relations among the closed-forms.
If no goals are passed, by default \Polar{} uses all (expected values
of) program variables of the (probabilistic) loop \texttt{BENCHMARK}.

%% file: 04-sensitivity.tex
\section{Parameters and Sensitivity Analysis}\label{sec:sensitivity}


\begin{figure}[t]
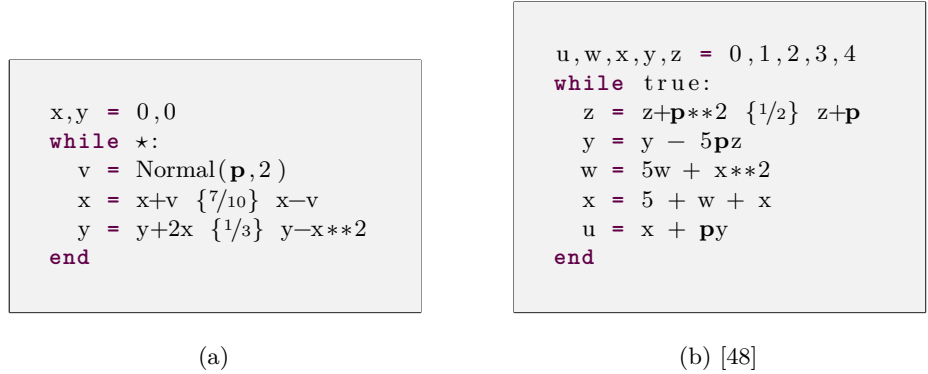

    \centering
    \begin{subfigure}[b]{0.45\textwidth}
        \begin{tcolorbox}[width=\linewidth,boxrule=0.5pt,arc=0pt,auto outer arc]
        \begin{lstlisting}
            x,y = 0,0
            while $\star$:
              v = $\text{Normal}$($\textbf{p}$,2)
              x = x+v {$\nicefrac{7}{10}$} x-v
              y = y+2x {$\nicefrac{1}{3}$} y-x**2
            end
        \end{lstlisting}
        \end{tcolorbox}
        \caption{}
        \label{fig:sensitivity:1}
    \end{subfigure}\hfill
    \begin{subfigure}[b]{0.45\textwidth}
        \begin{tcolorbox}[width=\linewidth,boxrule=0.5pt,arc=0pt,auto outer arc]
        \begin{lstlisting}
            u,w,x,y,z = 0,1,2,3,4
            while true:
              z = z+$\textbf{p}$**2 {$\nicefrac{1}{2}$} z+$\textbf{p}$
              y = y - 5$\textbf{p}$z
              w = 5w + x**2
              x = 5 + w + x
              u = x + $\textbf{p}$y
            end
        \end{lstlisting}
        \end{tcolorbox}
        \caption{\cite{MoosbruggerMK23}}
        \label{fig:sensitivity:2}
    \end{subfigure}
    \caption{Two probabilistic loops with computable sensitivities.}
    \label{fig:sensitivity}
\end{figure}

A major challenge when modeling stochastic processes with
probabilistic loops is that value distributions of some process
parameters are often unknown.
\Polar{} addresses this issue by supporting symbolic constants that represent arbitrary real numbers, enabling the modeling of such processes despite the unknown parameters.
In this context, the goal of \emph{sensitivity analysis} is to describe/quantify how small changes in these symbolic parameters affect the moments of loop variables.
As such, sensitivity analysis provides additional information about probabilistic loops when model parameters are only partially known.
The resulting sensitivity information can then be applied to various tasks, such as determining optimal values for unknown parameters based on given objectives.

\begin{example}\label{ex:sens:param}
Figure~\ref{fig:sensitivity:1} depicts a probabilistic loop with parameter $\texttt{p}$ representing an arbitrary real number.
Exploiting the techniques of Section~\ref{sec:closed-forms} and using symbolic constants, \Polar{} can compute the closed-forms of any moment of loop variables.
For instance, the closed-form of the expected value of $y$ is: 
\begin{equation*}
\E(y_n) = \frac{-8n^3\mathbf{p}^2 - 75n^2\mathbf{p}^2 + 30n^2\mathbf{p} - 150n^2 - 67n\mathbf{p}^2 + 30n\mathbf{p} - 150n}{225} 
\end{equation*}
Recall that the closed-forms computed by \Polar{} are exponential polynomials in the number of loop iterations $n$.
These closed-forms also include symbolic constants within the polynomials and exponential bases for parameterized loops.
Hence, the closed-forms for variable moments can be differentiated fully automatically with respect to a parameter leading to the sensitivities of variable moments. 
For the loop from Figure~\ref{fig:sensitivity:1}, we denote by $\diffp
\E(y_n)$ the sensitivity of the expected value of $y$ with respect to
$\texttt{p}$ and use \Polar{} to derive the following closed-form: 
\begin{equation*}
\diffp \E(y_n) = \frac{-16n^3\mathbf{p} - 150n^2\mathbf{p} + 30n^2 - 134n\mathbf{p} + 30n}{225}
\end{equation*}
\end{example}

\subsection{Sensitivity Analysis for Unsolvable Loops}\label{sec:sens:uns}

Analyzing sensitivities by utilizing closed-forms of variable moments requires that the loops satisfy the restrictions from Section~\ref{sec:closed-forms:restrictions}.
By providing an alternative method, \Polar{} is able to compute sensitivities even for loops that violate the restriction \emph{R3} of acyclic non-linearity.
For the loop in Figure~\ref{fig:sensitivity:2}, the program variable \texttt{w} depends \emph{non-linearly} on \texttt{x}.
Moreover, the variable \texttt{x} depends on \texttt{w}, creating a non-linear cyclic dependency.
This violates restriction \emph{R3}, and as a result, \Polar{} is unable to compute closed-forms for the moments of the variables.
In general, exponential polynomial closed-forms do not exist for such loops.
Hence, it is impossible to compute sensitivities by differentiating the closed-forms of variable moments.

With the technique of~\cite{MoosbruggerMK23},  the computation of
moments can be circumvented by constructing \emph{recurrences for
sensitivities}, instead of recurrences over moments, and utilizing the
potential independence of some variables and the parameter under
study. We illustrate this approach via the following example.  

\begin{example}
The variables \texttt{w} and \texttt{x} are responsible
for the violation of restriction \emph{R3} of the loop in
Figure~\ref{fig:sensitivity:2}; variables \texttt{w} and \texttt{x}
are however independent of the parameter \texttt{p}.
Therefore, changes in \texttt{p} do not affect moments that only involve \texttt{w} and \texttt{x}.
More precisely, $\diffp \E(M_n) = 0$ for any monomial $M_n$ in $w_n$ and $x_n$.
This fact can be utilized by constructing a recurrence for $\diffp \E(u_n)$:
\begin{flalign*}
    \diffp \E(u_n) &= \diffp \E(x_{n+1}) + \E(y_{n+1}) + \textbf{p} \diffp \E(y_{n+1}) \\
    &= \E(y_{n+1}) + \textbf{p} \diffp \E(y_{n+1})
\end{flalign*}
Using that $\diffp \E(x_n) = 0$, all variables responsible for the violation of restriction \emph{R3} vanish from the sensitivity recurrence of $\diffp \E(u_n)$.
The recurrence for $\diffp \E(u_n)$ can be completed to a system of C-finite recurrences by adding recurrences for $\E(y_n)$, $\diffp E(y_n)$, $\E(z_n)$, and $\diffp \E(z_n)$.
Hence, the sensitivity $\diffp \E(u_n)$ has an exponential polynomial closed-form even though the moment $\E(u_n)$ does not:
\begin{equation*}
\diffp \E(u_n) = -5n^2\mathbf{p}^3 - \frac{15}{4}n^2\mathbf{p}^2 - 5n\mathbf{p}^3 - \frac{15}{4}n\mathbf{p}^2 - 40n\mathbf{p} + 3
\end{equation*}
\end{example}

\subsection{Computing Parameter Sensitivities with \Polar} 

Computing sensitivities by differentiating the
closed-forms of moments, as detailed in Example~\ref{ex:sens:param}, is available in \Polar{} through its command-line interface:
\begin{Verbatim}[commandchars=\\\{\}]
    python polar.py BENCHMARK --goals GOALS \textbf{-sens_diff PARAM}
\end{Verbatim}
Using this command, \Polar{} computes the closed-forms of all goals
passed through the \texttt{goals} parameter and differentiates them
with respect to the parameter \texttt{PARAM}; for example, with
respect to parameter $\texttt{p}$ of Example~\ref{ex:sens:param}. 

Further, 
\Polar{} also 
derives recurrences for
sensitivities, as 
discussed in Section~\ref{sec:sens:uns}. To this end, \Polar{} uses simple static analysis techniques to determine parameter-independent program variables.
The functionality can be used by replacing the \texttt{sens_diff} parameter with \texttt{sens}:
\begin{Verbatim}[commandchars=\\\{\}]
    python polar.py BENCHMARK --goals GOALS \textbf{-sens PARAM}
\end{Verbatim}

%% file: 05-unsolvable-loops.tex
\section{Computational Hardness and Unsolvable Loops}\label{sec:unsolvable}


As argued in Section~\ref{sec:closed-forms}, the core
functionality of \Polar{} comes with computing closed-forms for (moments of)
loop variables using linear recurrences.
To ensure computability, \Polar{} imposes the three restrictions of
Section~\ref{sec:closed-forms:restrictions}
on its programming model. Lifting any of these restrictions brings
hard computational boundaries, as detailed next. 

Restriction \emph{R2} of
Section~\ref{sec:closed-forms:restrictions}, 
disallows arbitrary variables in if-conditions. Lifting this restriction would make \Polar{}'s programming model Turing-complete and render closed-forms uncomputable~\cite{polar}.
Additionally, recent results of~\cite{MullnerMK24} show that lifting
restriction \emph{R3}, which disallows arbitrary polynomial
dependencies within the loop body, render computing closed-forms and  the strongest
polynomial invariant \textsc{Skolem}-hard.
The \textsc{Skolem}-problem is a well-known problem in number theory
whose decidability has been unresolved for nearly a
century~\cite{everest2003recurrence,Tao-Skolem}. 
Furthermore, restriction \emph{R1} requires all probabilities to be constant.
Allowing for state-dependent probabilities would generally result in
cyclic non-linear dependencies in the moment recurrences, leading to
similar computational difficulties as encountered when lifting
restriction \emph{R3}.

Despite the restrictions of \emph{R1}--\emph{R3}, \Polar{} implements
additional techniques to analyze loops that cannot be solved by the
methods of Sections~\ref{sec:closed-forms}--\ref{sec:invariants}. We
refer to such loops as \emph{unsolvable loops}~\cite{unsolvable}.

\begin{figure}[t]
    \centering
    \begin{tcolorbox}[width=0.45\linewidth,boxrule=0.5pt,arc=0pt,auto outer arc]
    \begin{lstlisting}
        z = 0
        while $\star$:
            z = 1 - z
            x = 2x + y**2 + z
            y = 2y - y**2 + 2z
        end
    \end{lstlisting}
    \end{tcolorbox}
    \caption{An unsolvable loop \cite{unsolvableloopanalysis}.}
    \label{fig:unsolvable}
\end{figure}

\subsection{Closed-Forms for Unsolvable Loops}\label{sec:unsolvable:cf}

Intuitively, \emph{unsolvable} loops are loops with arbitrary
polynomial dependencies within their loop bodies. Hence, unsolvable
loops do not satisfy restriction \emph{R3}.
Despite the hardness results, \Polar{}  analyzes unsolvable loops
using incomplete, but sound methods.

\begin{example}
In the loop of Figure~\ref{fig:unsolvable}, the program variable
\texttt{y} depends non-linearly on itself, rendering the loop
unsolvable. Although exponential polynomial closed-forms for the variables \texttt{x} and \texttt{y} generally do not exist, combinations of these variables may allow for closed-forms.
For instance, the addition of \texttt{x} and \texttt{y} follows a linear recurrence and has the following closed-form:
\begin{equation}\label{eq:unsolv-cf}
x_n + y_n = 2^n (x_0 + y_0 + 2) - \frac{(-1)^n}{2} - \frac{3}{2}
\end{equation}
Intuitively, by summing up the variable \texttt{x} and \texttt{y}, the problematic non-linear self-dependency of \texttt{y} vanishes.
Generally, problematic non-linear dependencies may remain in linear
combinations of variables but vanish using polynomial combinations of
higher degrees.
\end{example}

\subsection{Solvable Loop Synthesis from Unsolvable Loops}\label{sec:unsolvable:synth}

Given an unsolvable loop, the polynomial combinations of program variables as described in Section~\ref{sec:unsolvable:cf} can be used to synthesize a solvable loop that overapproximates the original unsolvable loop~\cite{unsolvableloopanalysis}.

\begin{example}
The unsolvable loop from Figure~\ref{fig:unsolvable} is overapproximated by the following solvable loop with respect to the variable combination $x+y$.
\begin{center}
\begin{tcolorbox}[width=0.45\linewidth,boxrule=0.5pt,arc=0pt,auto outer arc]
\begin{lstlisting}
    s = $x_0 + y_0$
    while $\star$:
        t = 1 - z
        s = 2s + 3 - 3z
        z = t
    end
\end{lstlisting}
\end{tcolorbox}
\end{center}
The fresh program variable \texttt{s} simulates the value of the variable combination $x+y$.
Hence, \texttt{s} and $x+y$ share the closed-form stated in \eqref{eq:unsolv-cf}.
The solvable loop overapproximates the unsolvable loop from Figure~\ref{fig:unsolvable} in the sense that every invariant of the solvable loop is also an invariant of the unsolvable loop relative to the substitution $s \rightarrow x+y$.
\end{example}

\subsection{Analyzing Unsolvable Loops using \Polar}
In support of automating the analysis of unsolvable loops by computing
(unsolvable) closed-forms as discussed
in Section~\ref{sec:unsolvable:cf},
\Polar{} implements the techniques
of~\cite{unsolvable,unsolvableloopanalysis} to synthesize polynomial
combinations of program variables that obey linear recurrences,
whenever loop updates over individual variables may not yield linear
recurrences.
To synthesize such polynomial combinations of program variables for
unsolvable loops,  \Polar{} provides the following command-line interface:
\begin{Verbatim}[commandchars=\\\{\}]
    python polar.py BENCHMARK \textbf{--synth_unsolv_inv --inv_deg DEGREE}
\end{Verbatim}
Using the parameter \texttt{synth_unsolv_inv}, \Polar{} synthesizes
polynomial combinations of ``problematic'' program variables that follow a well-behaved recurrence~\cite{unsolvable}.
To circumvent the hardness results stemming from lifting (some)
restrictions of \emph{R1}--\emph{R3},
\Polar{} synthesizes all such polynomial combinations, where the degree of the polynomials is bounded by the natural number \texttt{DEGREE} passed through the \texttt{inv_deg} parameter.
For probabilistic input loops, the closed-forms are for the expected value of the synthesized combinations instead of their exact values.

Further, 
\Polar{} implements the approach of~\cite{unsolvableloopanalysis} to
synthesize solvable loops from unsolvable loops, as discussed in
Section~\ref{sec:unsolvable:synth}. In essence, \Polar{}'s 
synthesis procedure replaces the program variables ``responsible for
unsolvability'' with a fresh variable that simulates a well-behaved polynomial combination of program variables, as introduced in Section~\ref{sec:unsolvable:cf}.
The solvable loop synthesized by \Polar{} is always deterministic, even if the original unsolvable loop is probabilistic.
In the case of a probabilistic input loop, the variables in the
synthesized loop represent the expected values of the original
variables and the combination of program variables. 
To synthesize solvable loops from a given unsolvable loop \Polar{} provides the following command-line interface:
\begin{Verbatim}[commandchars=\\\{\}]
    python polar.py BENCHMARK \textbf{--synth_solv_loop} --inv_deg DEGREE
\end{Verbatim}
By calling this command, \Polar{} first computes all well-behaved polynomial combinations of program variables up to the fixed degree \texttt{DEGREE}.
For every combination found, \Polar{} synthesizes a solvable loop that
overapproximates the original unsolvable loop of \texttt{BENCHMARK}. 

%% file: 06-related-work.tex
\section{Related Work}

While the verification and invariant generation of (probabilistic) loops are widely studied research problems with many theoretical challenges, our focus with \Polar{} remains in the computer-aided efficient automation of loop analysis.

For classical programs, recurrence relations serve as common workhorses for program analysis~\cite{Kovacs2008,Humenberger2018,Farzan2015,KincaidCBR18}.
Inferring polynomial invariants by using linear recurrences and Gr\"obner bases computation was pioneered in \cite{Rodriguez-carbonell2004,RCarbonellK07} and later extended to more general recurrences \cite{Kovacs2008,Humenberger2018}.
The tool \textsc{Aligator}~\cite{Aligator18} can automatically compute the strongest polynomial invariant for a class of \emph{classical}  loops.
However, the loops supported by \textsc{Aligator} are a strict subset of those supported by \Polar{}, as \textsc{Aligator} does not handle any form of randomness or branching statements.
The use of linear recurrences for probabilistic loops was first introduced in~\cite{Bartocci2019}, leading to the development of the tool \textsc{Mora}~\cite{Bartocci2020a}.
While \textsc{Mora} can compute closed-forms for a strict subset of probabilistic loops supported by \Polar{}, it lacks additional functionalities such as sensitivity analysis and invariant synthesis.

Due to the uncomputability of invariants for general probabilistic programs, many techniques resort to incomplete techniques.
Constraint-based methods use templates for potential invariants and aim to solve for template instantiations resulting in invariants \cite{FengZJZX17,KatoenMMM10}.
In \cite{Bao2021} the authors use neural networks as potential invariants and train them using counter-examples.
Another technique leveraging counter-examples to guide the search for probabilistic invariants was introduced in \cite{BatzCJKKM23}.
In some cases, probabilistic programs can be automatically verified without synthesizing invariants by resorting to induction techniques \cite{Batz2021}.
Alternatively to invariant synthesis, user-provided annotations can be used within a deductive verification framework \cite{SchroerBKKM23}.

Termination is a property not yet addressed by \Polar{}. 
Most tools for probabilistic termination analysis use martingale-based proof rules, automated through constraint solving \cite{chakarov2013probabilistic,agrawal-lexicographic-2017,ChenH20,ChatterjeeNZ17,chatterjee-termination-2016}.
Alternatively, the automation techniques implemented in the probabilistic termination tool \textsc{Amber}~\cite{moosbruggerFM,MoosbruggerBKK22} are based on recurrence relations~\cite{moosbrugger2020automated}.
Analogous to invariant synthesis, neural networks, and learning techniques have been employed for probabilistic termination analysis \cite{AbateGR20}.
Additionally, tools for expected cost analysis can also be used to verify finite expected runtimes \cite{ngo-bounded-2018,meyer2021inferring,avanzini2020modular}. Extending \Polar{} with automated methods (dis)proving loop termination is an interesting direction for future work. 

%% file: 07-conclusion.tex
\section{Conclusion}
We present the \Polar{} framework for the automated analysis of classical and probabilistic loops using algebraic recurrences.
\Polar{} supports the analysis of (probabilistic) loops with if-statements, polynomial arithmetic, and draws from common probability distributions.
The core functionality of \Polar{} is the extraction of C-finite recurrences that describe the moments of loop variables.
These recurrences yield exponential polynomial closed-forms, which \Polar{}  uses to compute the strongest polynomial invariant of a loop fully automatically.

To ensure computability for the class of supported input programs, \Polar{} imposes certain restrictions on the variables within branching statements and allows only acyclic polynomial dependencies.
In addition to closed-forms and invariants, \Polar{}  can also be used to compute the sensitivities of (moments of) program variables with respect to unknown model parameters.
For loops violating the restriction of acyclic polynomial dependencies, \Polar{} offers an incomplete, but sound,  method for inferring closed-forms of combinations of variables.
As such, \Polar{} provides a multitude of algebraic features for the exact and automated analysis of (probabilistic) loops.

%% file: references.bib
@inproceedings{AbateGR20,
  author       = {Alessandro Abate and
                  Mirco Giacobbe and
                  Diptarko Roy},
  title        = {Learning Probabilistic Termination Proofs},
  booktitle    = {Proc. of {CAV}},
  year         = {2021},
  doi          = {10.1007/978-3-030-81688-9\_1},
}

@inproceedings{KatoenMMM10,
  author       = {Joost{-}Pieter Katoen and
                  Annabelle McIver and
                  Larissa Meinicke and
                  Carroll C. Morgan},
  title        = {Linear-Invariant Generation for Probabilistic Programs: - Automated
                  Support for Proof-Based Methods},
  booktitle    = {Proc. of {SAS}},
  year         = {2010},
  doi          = {10.1007/978-3-642-15769-1\_24},
}

@inproceedings{FengZJZX17,
  author       = {Yijun Feng and
                  Lijun Zhang and
                  David N. Jansen and
                  Naijun Zhan and
                  Bican Xia},
  title        = {Finding Polynomial Loop Invariants for Probabilistic Programs},
  booktitle    = {Proc. of {ATVA}},
  year         = {2017},
  doi          = {10.1007/978-3-319-68167-2\_26},
}

@article{SchroerBKKM23,
  author       = {Philipp Schr{\"{o}}er and
                  Kevin Batz and
                  Benjamin Lucien Kaminski and
                  Joost{-}Pieter Katoen and
                  Christoph Matheja},
  title        = {A Deductive Verification Infrastructure for Probabilistic Programs},
  journal      = {Proc. {ACM} Program. Lang.},
  number       = {{OOPSLA2}},
  year         = {2023},
  doi          = {10.1145/3622870},
}

@inproceedings{KlinkenbergBKKM20,
  author       = {Lutz Klinkenberg and
                  Kevin Batz and
                  Benjamin Lucien Kaminski and
                  Joost{-}Pieter Katoen and
                  Joshua Moerman and
                  Tobias Winkler},
  editor       = {Maribel Fern{\'{a}}ndez},
  title        = {Proc. of {LOPSTR}},
  year         = {2020},
  doi          = {10.1007/978-3-030-68446-4\_12},
}

@Article{Kozen1985,
  author  = {Dexter Kozen},
  title   = {A Probabilistic {PDL}},
  journal = {J. Comput. Syst. Sci.},
  year    = {1985},
  doi     = {10.1016/0022-0000(85)90012-1},
}

@book{Barthe2020,
  author =        {Gilles Barthe and Joost-Pieter Katoen and
                   Alexandra Silva},
  publisher =     {Cambridge University Press},
  title =         {Foundations of Probabilistic Programming},
  year =          {2020},
  doi =           {10.1017/9781108770750},
}

@article{Ghahramani2015,
  author =        {Zoubin Ghahramani},
  journal =       {Nature},
  title =         {Probabilistic Machine Learning and Artificial
                   Intelligence},
  year =          {2015},
  doi =           {10.1038/nature14541},
}

@inproceedings{Kaminski2016,
  author =        {Benjamin Lucien Kaminski and Joost{-}Pieter Katoen and
                   Christoph Matheja and Federico Olmedo},
  booktitle =     {Proc. of ESOP},
  title =         {Weakest Precondition Reasoning for Expected Run-Times
                   of Probabilistic Programs},
  year =          {2016},
  doi =           {10.1007/978-3-662-49498-1},
}

@inproceedings{Selyunin2015,
  author =        {Konstantin Selyunin and Denise Ratasich and
                   Ezio Bartocci and Md. Ariful Islam and
                   Scott A. Smolka and Radu Grosu},
  booktitle =     {Proc. of CDC},
  title =         {Neural Programming: Towards Adaptive Control in
                   Cyber-Physical Systems},
  year =          {2015},
  doi =           {10.1109/CDC.2015.7403319},
}

@inproceedings{Chou2020,
  author =        {Yi Chou and Hansol Yoon and Sriram Sankaranarayanan},
  booktitle =     {Proc. of IROS},
  title =         {Predictive Runtime Monitoring of Vehicle Models Using
                   Bayesian Estimation and Reachability Analysis},
  year =          {2020},
  doi =           {10.1109/IROS45743.2020.9340755},
}

@inproceedings{Barthe2012,
  author =        {Gilles Barthe and Benjamin Gr{\'{e}}goire and
                   Santiago Zanella B{\'{e}}guelin},
  booktitle =     {Proc. of {MPC}},
  title =         {Probabilistic Relational Hoare Logics for
                   Computer-Aided Security Proofs},
  year =          {2012},
  doi =           {10.1007/978-3-642-31113-0},
}

@inproceedings{Barthe2012a,
  author =        {Gilles Barthe and Boris K{\"{o}}pf and
                   Federico Olmedo and Santiago Zanella B{\'{e}}guelin},
  booktitle =     {Proc. of {POPL}},
  title =         {Probabilistic Relational Reasoning for Differential
                   Privacy},
  year =          {2012},
  doi =           {10.1145/2103656.2103670},
}

@book{Motwani1995,
  author =        {Rajeev Motwani and Prabhakar Raghavan},
  publisher =     {Cambridge University Press},
  title =         {Randomized Algorithms},
  year =          {1995},
  doi =           {10.1017/cbo9780511814075},
}

@book{McIver2005,
  author =        {Annabelle McIver and Carroll Morgan},
  publisher =     {Springer},
  title =         {Abstraction, Refinement and Proof for Probabilistic
                   Systems},
  year =          {2005},
  doi =           {10.1007/b138392},
}

@inproceedings{Bartocci2020a,
  author =        {Ezio Bartocci and Laura Kov{\'{a}}cs and
                   Miroslav Stankovic},
  booktitle =     {Proc. of {TACAS}},
  title =         {Mora - Automatic Generation of Moment-Based
                   Invariants},
  year =          {2020},
  doi =           {10.1007/978-3-030-45190-5\_28},
}

@article{Stankovic2022,
  author =        {Miroslav Stankovic and Ezio Bartocci and
                   Laura Kov{\'{a}}cs},
  journal =       {Theor. Comput. Sci.},
  title =         {Moment-Based Analysis of Bayesian Network Properties},
  year =          {2022},
  doi =           {10.1016/j.tcs.2021.12.021},
}

@book{Kauers2011,
  author =        {Manuel Kauers and Peter Paule},
  publisher =     {Springer},
  title =         {The Concrete Tetrahedron - Symbolic Sums, Recurrence
                   Equations, Generating Functions, Asymptotic
                   Estimates},
  year =          {2011},
  doi =           {10.1007/978-3-7091-0445-3},
}

@inproceedings{Bartocci2019,
  author =        {Ezio Bartocci and Laura Kov{\'{a}}cs and
                   Miroslav Stankovic},
  booktitle =     {Proc. of {ATVA}},
  title =         {Automatic Generation of Moment-Based Invariants for
                   Prob-Solvable Loops},
  year =          {2019},
  doi =           {10.1007/978-3-030-31784-3\_15},
}

@inproceedings{Batz2021,
  author =        {Batz, Kevin and Chen, Mingshuai and
                   Kaminski, Benjamin Lucien and Katoen, Joost-Pieter and
                   Matheja, Christoph and Schr{\"o}er, Philipp},
  booktitle =     {Proc. of {CAV}},
  title =         {Latticed k-Induction with an Application to
                   Probabilistic Programs},
  year =          {2021},
  doi =           {10.1007/978-3-030-81688-9_25},
}

@inproceedings{Farzan2015,
  author =        {Farzan, Azadeh and Kincaid, Zachary},
  booktitle =     {Proc. of FMCAD},
  title =         {Compositional Recurrence Analysis},
  year =          {2015},
  doi =           {10.1109/FMCAD.2015.7542253},
}

@inproceedings{Humenberger2018,
  author =        {Andreas Humenberger and Maximilian Jaroschek and
                   Laura Kov{\'a}cs},
  booktitle =     {Proc. of VMCAI},
  title =         {Invariant Generation for Multi-Path Loops with
                   Polynomial Assignments},
  year =          {2018},
  doi =           {10.1007/978-3-319-73721-8_11},
}

@inproceedings{Kovacs2008,
  author =        {Laura Kov{\'a}cs},
  booktitle =     {Proc. of TACAS},
  title =         {Reasoning Algebraically About P-Solvable Loops},
  year =          {2008},
  doi =           {10.1007/978-3-540-78800-3_18},
}

@inproceedings{Rodriguez-carbonell2004,
  author =        {Enric Rodr{\'i}guez-carbonell and Deepak Kapur},
  booktitle =     {Proc. of ISSAC},
  title =         {Automatic Generation of Polynomial Loop Invariants:
                   Algebraic Foundations},
  year =          {2004},
  doi =           {10.1145/1005285.1005324},
}

@article{Katoen2011,
  author =        {Joost{-}Pieter Katoen and Ivan S. Zapreev and
                   Ernst Moritz Hahn and Holger Hermanns and
                   David N. Jansen},
  journal =       {Perform. Eval.},
  title =         {The Ins and Outs of the Probabilistic Model Checker
                   {MRMC}},
  year =          {2011},
  doi =           {10.1016/j.peva.2010.04.001},
}

@inproceedings{Bao2021,
  author =        {Jialu Bao and Nitesh Trivedi and Drashti Pathak and
                   Justin Hsu and Subhajit Roy},
  booktitle =     {Proc. of {CAV}},
  title =         {Data-Driven Invariant Learning for Probabilistic
                   Programs},
  year =          {2022},
  doi =           {10.1007/978-3-031-13185-1\_3},
}

@InProceedings{unsolvable,
  author    = {Daneshvar Amrollahi and Ezio Bartocci and George Kenison and Laura Kov{\'{a}}cs and Marcel Moosbrugger and Miroslav Stankovic},
  title     = {{Solving Invariant Generation for Unsolvable Loops}},
  booktitle = {Proc. of {SAS}},
  year      = {2022},
  doi       = {10.1007/978-3-031-22308-2\_3},
}

@Article{Aguirre2021,
  author  = {Alejandro Aguirre and Gilles Barthe and Justin Hsu and Benjamin Lucien Kaminski and Joost{-}Pieter Katoen and Christoph Matheja},
  title   = {A pre-expectation calculus for probabilistic sensitivity},
  journal = {Proc. {ACM} Program. Lang.},
  year    = {2021},
  number  = {{POPL}},
  doi     = {10.1145/3434333},
}

@Article{polar,
  author  = {Marcel Moosbrugger and Miroslav Stankovic and Ezio Bartocci and Laura Kov{\'{a}}cs},
  title   = {{This Is The Moment for Probabilistic Loops}},
  journal = {Proc. {ACM} Program. Lang.},
  year    = {2022},
  number  = {{OOPSLA2}},
  doi     = {10.1145/3563341},
}

@Article{RCarbonellK07,
  author  = {Enric Rodr{\'{\i}}guez{-}Carbonell and Deepak Kapur},
  title   = {{Generating All Polynomial Invariants in Simple Loops}},
  journal = {J. Symb. Comput.},
  year    = {2007},
  doi     = {10.1016/j.jsc.2007.01.002},
}

@Book{Tao-Skolem,
  title     = {Structure and Randomness},
  publisher = {American Mathematical Society},
  year      = {2008},
  author    = {Terrence Tao},
  note      = {ISBN 0-8218-4695-7},
}

@InProceedings{KofnovMSBB22,
  author    = {Andrey Kofnov and Marcel Moosbrugger and Miroslav Stankovic and Ezio Bartocci and Efstathia Bura},
  title     = {Moment-Based Invariants for Probabilistic Loops with Non-polynomial Assignments},
  booktitle = {Proc. of {QEST}},
  year      = {2022},
  doi       = {10.1007/978-3-031-16336-4\_1},
}

@Book{CoxLittleOshea97,
  title     = {Ideals, varieties, and algorithms - an introduction to computational algebraic geometry and commutative algebra {(2.} ed.)},
  publisher = {Springer},
  year      = {1997},
  author    = {David A. Cox and John Little and Donal O'Shea},
  isbn      = {978-0-387-94680-1},
  doi       = {10.1137/1035171},
}

@Article{Buchberger-thesis,
  author  = {Bruno Buchberger},
  title   = {Bruno Buchberger's PhD thesis 1965: An algorithm for finding the basis elements of the residue class ring of a zero dimensional polynomial ideal},
  journal = {J. Symb. Comput.},
  year    = {2006},
  doi     = {10.1016/j.jsc.2005.09.007},
}

@PhdThesis{kauers-thesis,
  author = {Manuel Kauers},
  title  = {Algorithms for Nonlinear Higher Order Difference Equations},
  school = {RISC, Johannes Kepler University, Linz},
  year   = {2005},
}

@inproceedings{KauersNP23,
  author       = {Manuel Kauers and
                  Philipp Nuspl and
                  Veronika Pillwein},
  title        = {Order bounds for C2-finite sequences},
  booktitle    = {Proc. of {ISSAC}},
  year         = {2023},
  doi          = {10.1145/3597066.3597070},
}

@Article{KincaidCBR18,
  author  = {Zachary Kincaid and John Cyphert and Jason Breck and Thomas W. Reps},
  title   = {Non-linear reasoning for invariant synthesis},
  journal = {Proc. {ACM} Program. Lang.},
  year    = {2018},
  number  = {{POPL}},
  doi     = {10.1145/3158142},
}

@InProceedings{BatzCJKKM23,
  author    = {Kevin Batz and Mingshuai Chen and Sebastian Junges and Benjamin Lucien Kaminski and Joost{-}Pieter Katoen and Christoph Matheja},
  title     = {Probabilistic Program Verification via Inductive Synthesis of Inductive Invariants},
  booktitle = {Proc. of {TACAS}},
  year      = {2023},
  doi       = {10.1007/978-3-031-30820-8\_25},
}

@Article{KaminskiKMO18,
  author  = {Benjamin Lucien Kaminski and Joost{-}Pieter Katoen and Christoph Matheja and Federico Olmedo},
  title   = {Weakest Precondition Reasoning for Expected Runtimes of Randomized Algorithms},
  journal = {J. {ACM}},
  year    = {2018},
  doi     = {10.1145/3208102},
}

@Article{BatzKKMV23,
  author  = {Kevin Batz and Benjamin Lucien Kaminski and Joost{-}Pieter Katoen and Christoph Matheja and Lena Verscht},
  title   = {A Calculus for Amortized Expected Runtimes},
  journal = {Proc. {ACM} Program. Lang.},
  year    = {2023},
  number  = {{POPL}},
  doi     = {10.1145/3571260},
}

@InProceedings{ChatterjeeNZ17,
  author    = {Krishnendu Chatterjee and Petr Novotn{\'{y}} and Dorde Zikelic},
  title     = {Stochastic invariants for probabilistic termination},
  booktitle = {Proc. of {POPL}},
  year      = {2017},
  doi       = {10.1145/3009837.3009873},
}

@InProceedings{chakarov2013probabilistic,
  author    = {Aleksandar Chakarov and Sriram Sankaranarayanan},
  title     = {Probabilistic Program Analysis with Martingales},
  booktitle = {Proc. of {CAV}},
  year      = {2013},
  doi       = {10.1007/978-3-642-39799-8\_34},
}

@InProceedings{Aligator18,
  author    = {Andreas Humenberger and Maximilian Jaroschek and Laura Kov{\'{a}}cs},
  title     = {Aligator.jl - {A} Julia Package for Loop Invariant Generation},
  booktitle = {Proc. of {CICM}},
  year      = {2018},
  doi       = {10.1007/978-3-319-96812-4\_10},
}

@InProceedings{ChenH20,
  author    = {Jianhui Chen and Fei He},
  title     = {Proving almost-sure termination by omega-regular decomposition},
  booktitle = {Proc. of {PLDI}},
  year      = {2020},
  doi       = {10.1145/3385412.3386002},
}

@InProceedings{ngo-bounded-2018,
  author    = {Ngo, Van Chan and Carbonneaux, Quentin and Hoffmann, Jan},
  title     = {Bounded expectations: resource analysis for probabilistic programs},
  booktitle = {Proc. of PLDI},
  year      = {2018},
  doi       = {10.1145/3192366.3192394},
}

@Article{hark-aiming-2020,
  author  = {Marcel Hark and Benjamin Lucien Kaminski and J{\"{u}}rgen Giesl and Joost{-}Pieter Katoen},
  title   = {Aiming low is harder: induction for lower bounds in probabilistic program verification},
  journal = {Proc. {ACM} Program. Lang.},
  year    = {2020},
  number  = {{POPL}},
  doi     = {10.1145/3371105},
}

@InProceedings{chatterjee-termination-2016,
  author    = {Chatterjee, Krishnendu and Fu, Hongfei and Goharshady, Amir Kafshdar},
  title     = {Termination {Analysis} of {Probabilistic} {Programs} {Through} {Positivstellensatz}’s},
  booktitle = {Proc. of CAV},
  year      = {2016},
  doi       = {10.1007/978-3-319-41528-4\_1},
}

@Article{agrawal-lexicographic-2017,
  author  = {Sheshansh Agrawal and Krishnendu Chatterjee and Petr Novotn{\'{y}}},
  title   = {Lexicographic ranking supermartingales: an efficient approach to termination of probabilistic programs},
  journal = {Proc. {ACM} Program. Lang.},
  year    = {2018},
  number  = {{POPL}},
  doi     = {10.1145/3158122},
}

@InProceedings{meyer2021inferring,
  author        = {Fabian Meyer and Marcel Hark and J{\"{u}}rgen Giesl},
  title         = {{Inferring Expected Runtimes of Probabilistic Integer Programs Using Expected Sizes}},
  booktitle     = {Proc. of {TACAS}},
  year          = {2021},
  __markedentry = {[marcel:]},
  doi           = {10.1007/978-3-030-72016-2\_14},
}

@Article{avanzini2020modular,
  author        = {Martin Avanzini and Georg Moser and Michael Schaper},
  title         = {A modular cost analysis for probabilistic programs},
  journal       = {Proc. {ACM} Program. Lang.},
  year          = {2020},
  number        = {{OOPSLA}},
  __markedentry = {[marcel:]},
  doi           = {10.1145/3428240},
}

@InProceedings{moosbrugger2020automated,
  author        = {Marcel Moosbrugger and Ezio Bartocci and Joost-Pieter Katoen and Laura Kov{\'{a}}cs},
  title         = {{Automated Termination Analysis of Polynomial Probabilistic Programs}},
  booktitle     = {Proc. of {ESOP}},
  year          = {2021},
  __markedentry = {[marcel:]},
  doi           = {10.1007/978-3-030-72019-3\_18},
}

@InProceedings{moosbruggerFM,
  author        = {Marcel Moosbrugger and Ezio Bartocci and Joost-Pieter Katoen and Laura Kov{\'{a}}cs},
  title         = {{The Probabilistic Termination Tool Amber}},
  booktitle     = {Proc. of {FM}},
  year          = {2021},
  __markedentry = {[marcel:]},
  doi           = {10.1007/978-3-030-90870-6\_36},
}

@Book{everest2003recurrence,
  title         = {Recurrence Sequences},
  publisher     = {Amer. Math. Soc.},
  year          = {2003},
  author        = {Graham Everest and Alfred J. van der Poorten and Igor E. Shparlinski and Thomas Ward},
  series        = {Math. Surveys Monogr.},
  address       = {Providence},
  isbn          = {978-0-8218-3387-2},
  note          = {ISBN 978-0-8218-3387-2},
  __markedentry = {[marcel:]},
}

@InProceedings{MoosbruggerMK23,
  author    = {Marcel Moosbrugger and Julian M{\"{u}}llner and Laura Kov{\'{a}}cs},
  title     = {{Automated Sensitivity Analysis for Probabilistic Loops}},
  booktitle = {Proc. of {iFM}},
  year      = {2023},
  doi       = {10.1007/978-3-031-47705-8\_2},
}

@Article{MullnerMK24,
  author  = {Julian M{\"{u}}llner and Marcel Moosbrugger and Laura Kov{\'{a}}cs},
  title   = {{Strong Invariants Are Hard: On the Hardness of Strongest Polynomial Invariants for (Probabilistic) Programs}},
  journal = {Proc. {ACM} Program. Lang.},
  year    = {2024},
  number  = {{POPL}},
  doi     = {10.1145/3632872},
}

@Article{unsolvableloopanalysis,
  author  = {Daneshvar Amrollahi and Ezio Bartocci and George Kenison and Laura Kov{\'{a}}cs and Marcel Moosbrugger and Miroslav Stankovic},
  title   = {{(Un)Solvable Loop Analysis}},
  journal = {Formal Methods Syst. Des.},
  year    = {2024},
  note    = {{To appear.}},
}

@Article{MoosbruggerBKK22,
  author  = {Marcel Moosbrugger and Ezio Bartocci and Joost{-}Pieter Katoen and Laura Kov{\'{a}}cs},
  title   = {{The Probabilistic Termination Tool Amber}},
  journal = {Formal Methods Syst. Des.},
  year    = {2022},
  doi     = {10.1007/S10703-023-00424-Z},
}

@Article{Kofnov2024,
  author  = {Kofnov, Andrey and Moosbrugger, Marcel and Stankovi\v{c}, Miroslav and Bartocci, Ezio and Bura, Efstathia},
  title   = {Exact and Approximate Moment Derivation for Probabilistic Loops With Non-Polynomial Assignments},
  journal = {ACM Trans. Model. Comput. Simul.},
  year    = {2024},
  note    = {Just Accepted},
  doi     = {10.1145/3641545},
}
